\documentclass[aps,pra,twocolumn,showpacs,floatfix]{revtex4}
\usepackage{bm}
\usepackage{graphicx} 
\usepackage{amsmath}
\begin{document}
\title{Long-Lived Complexes and Chaos in Ultracold Molecular Collisions}
\author{James F. E. Croft}
\author{John L. Bohn}
\affiliation{JILA, NIST, and Department of Physics, University of Colorado, Boulder, Colorado 80309-0440, USA}

\date{\today}

\begin{abstract}
Estimates for the lifetime of collision complexes formed during ultracold molecular collisions based on density-of-states arguments are shown to be consistent with similar estimates based on classical trajectory calculations.  In the classical version, these collisions are shown to exhibit chaos, and their fractal dimension is calculated versus collision energy.  From these results, a picture emerges that ultracold collisions are likely classically ergodic, justifying the density-of-states estimates for lifetimes.  These results point the way toward using the techniques of classical and quantum chaos to interpret molecular collisions in the ultracold regime.  
\end{abstract}


\maketitle
\section{Introduction}
\label{intro}
Ultracold gases provide a unique environment for molecules, where translational temperature can be far less than the interaction energy of molecules with one another.  This is by now a familiar circumstance in ultracold physics and one that affords among other things the creation of novel quantum states of matter \cite{Carr:NJPintro:2009,Wall:2009} and the sensitive control over chemical kinematics \cite{Bell:2009,Ospelkaus:react:2010}. These effects typically rely on the relatively large strength of long-range interactions between the molecules, e.g., dipole-dipole forces.

By contrast, recent work has postulated that ultralow temperatures may also influence and probe the detailed interaction between molecules at short-range, in the act of colliding, in particular by vastly increasing the interaction time between the molecules.  This effect arises from the huge disparity between the energy scales of the intermolecular potential energy surface ($\sim 10^3$ K), and the translational temperature of the free molecules ($\ll 10^{-3}$ K).  Upon colliding, the molecules accelerate into the potential well, converting this potential energy into internal rotational and vibrational motions of the molecules.  Time spent in these modes of the molecular ``complex'' contribute to long dwell times within the complex before it fragments into free molecules, thus probing large portions of the potential energy surface.

At the very simplest level of understanding, this idea is codified in the Rice-Ramsperger-Kassel-Marcus (RRKM) approximation \cite{Marcus:Vol1:1952,Marcus:Vol2:1952,Levine:molecular:2005}, where the dwell time of the complex is approximated as
\begin{eqnarray}
\label{eqn:rrkm}
\tau_{dos} = \frac{ 2 \pi  \hbar \rho }{ N_o }.
\end{eqnarray}
Here $\rho$ is the density of available ro-vibrational states (DOS), while $N_o$ is the number of open channels, i.e., quantum states energetically available to the collision fragments.  This formula expresses the simple idea that the atoms comprising the complex have many ways of distributing their energy (as counted by $\rho$) so as {\it not} to fragment back into separate molecules. They thus spend a long time exploring phase space before finding one of the comparatively small number of ways $N_o$ to fragment.  At room temperature $N_o$ may be large, consequently the complex's lifetime may be negligible.  However, in the ultracold environment, the value of $N_o$ can plummet all the way to $N_o=1$, meaning that the complex must restore the molecules exactly to their initial quantum states before fragmentation can occur.  It is this circumstance -- small $N_o$ -- that is novel in the ultracold environment.  

Mayle {\it et al} \cite{Mayle:2012,Mayle:2013} used this idea as a point of departure, from the usual quantum scattering methods used in ultracold physics, to assess the behavior of the collision complex.  The lifetime of the complex was indeed found to be long, of order 10-100 ns for alkali atom-alkali dimer collisions, and of order 1-10 msec for collisions of alkali molecules with one another.  Because the latter timescale is comparable to experimental lifetimes, the existence of complexes may lead to novel trap loss mechanisms, such as described in Ref.~\cite{Mayle:2013}.  In addition, Refs \cite{Mayle:2012,Mayle:2013}  assessed aspects of spin dynamics in the complex, statistics of resonant energy-level spacing, and statistical aspects of scattering such as Ericson fluctuations \cite{Ericson:1960,Ericson:1963}.  

The key feature that makes RRKM theory work is the assertion that all states contributing to $\rho$ actually get explored during the typical collision, so that the estimate of time wasted is accurate.  This is not necessarily the case, as for example when the incident molecules are separated from much of phase space by barriers in the potential energy surface, or else when the number of open channels $N_o$ is so large that a typical trajectory leaves before seeing all the states available \cite{Oref:1979,Diau:1998}. In ultracold collisions of alkali molecules, potential energy surfaces  are likely to be barrierless \cite{Tscherbul:Rb2Cs:2008,Zemke:2010,Byrd:2010,Byrd:thesis:2013}, whereby the full DOS should be accessible.

In this paper we provide theoretical evidence that the lifetimes based on RRKM estimates agree to within an order of magnitude with the results of classical trajectory calculations that yield explicit dwell times. We interpret these results to mean that collisions in this regime are ergodic, consistent with the foundations of RRKM theory. Moreover, the lifetimes of various trajectories are found to be extremely sensitive functions of initial conditions, illustrating that classically chaotic dynamics is at work.  We quantify the onset of chaos in terms of a ``fractal dimension'' for the space of incident conditions, finding that classical chaos emerges well above ultracold energies.

\section{Classical Trajectory Calculations}
\label{sec:theo}

Viewed as a problem in multichannel quantum mechanics, the presence of a vast number of ro-vibrational resonant states would necessitate an equally vast set of scattering channels, rendering the problem extremely difficult, if not impossible.  The explicit consideration of nuclear spin would of course make this problem even worse. In addition for collisions in an applied field, the total angular momentum $J$ is no longer a good quantum number and the large sets of coupled equations can no longer be factorized neatly into smaller blocks for each $J$ as is possible in field-free scattering.

Even if such calculations were easily done, they would still likely not yield accurate resonance positions, since these are extremely sensitive to the  potential surfaces.  These surfaces are themselves computationally intensive and are thus often only accurate to a couple of percent. For cold atomic collisions the potential has to be modified in order to fit experimental observables \cite{Takekoshi:RbCs:2012,Ruzic:2013} and for molecular collisions it is necessary to vary the potential by a factor and to content oneself with the study of general trends \cite{Wallis:MgNH:2009}.

While quantitative work is in progress to mitigate the expense of such computations \cite{TVTscherbul:total:2010,Croft:MQDT:2011}, quantitative level assignment of resonance lines seems a distant goal.  In this regime observables become averaged over many resonances and taking a statistical approach to cold collisions such as the lifetime in Equation (\ref{eqn:rrkm}) is apposite \cite{Flambaum:2006,Mayle:2012,Mayle:2013}.

Here we take an alternative, time-honored approach, and estimate the overall properties of ultracold alkali molecule collisions via classical trajectory simulations.  Doing so it is fairly straightforward to extract mean lifetimes from an ensemble of trajectories using a topologically reasonable, approximate  potential energy surface (PES).  In this section we describe our approach and the PES used.

\subsection{Classical Trajectory calculations and Initial Conditions}
\begin{figure}[tbp]
\centering
\includegraphics[width=1.0\columnwidth]{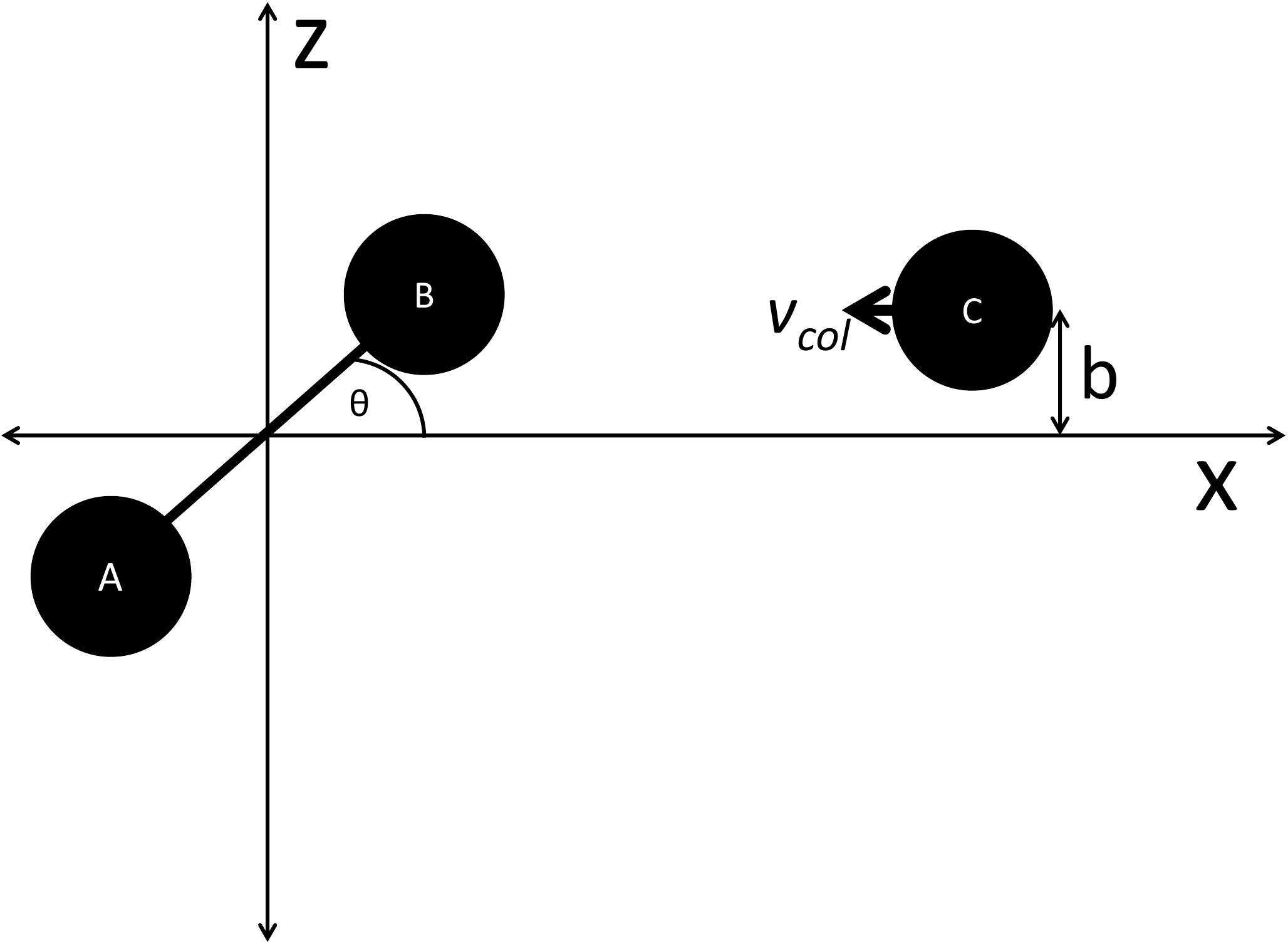}
\caption{Schematic of the initial conditions. The lone atom is given an initial velocity $v_{col}$ corresponding to a collision energy $E_{col}$. For collisions with an impact parameter the lone atom is displaced a distance $b$ in the $yz$ plane.} 
\label{fig:initial_set_up}
\end{figure}

The collision calculations are performed in the coordinate system depicted in Figure \ref{fig:initial_set_up}. We start by placing the center of mass of the diatom (atoms A and B) at the origin along the $x$ axis in the $xz$ plane, with the atoms at the equilibrium bond length with zero momentum. For calculations with zero impact parameter the lone atom (labeled C) is then placed on the $x$ axis at a distance $R_\infty$ (where $R_\infty$ is sufficiently large that the lone atom is effectively moving  freely) and the dimer set to an angle $\theta$ relative to the lone atom. The lone atom is then given an initial kinetic energy of $E_{col}$ relative to the origin. For collisions with nonzero impact parameter the lone atom is further displaced in the $yz$ plane such that the total of angular momentum is equal to a given choice $l$.

All classical trajectory calculations were performed using The Adiabatic and Nonadiabatic Trajectories (ANT) package \cite{ANT:2013}. The 4th order Runge-Kutta method with fixed step size was used to propagate Hamilton's equations of motion. Trajectories were considered complete when a lone atom had been ejected and was again a distance $R_\infty$ from the dimer with enough kinetic energy to escape the potential of the dimer.

All computed trajectories are necessarily done to a finite precision. However since chaotic systems display sensitive dependence on initial conditions, computed trajectories diverge exponentially from the true trajectory with the same initial conditions. In this work we assume shadowing, that is, that there exists an errorless trajectory with a slightly different initial condition that shadows the computed one \cite{Hammel:Shadow:1987,Greborgi:Shadow:1990}. Properties averaged over a large number of trajectories thus yield a meaningful result. 

Ordinarily one expects classical approximations to be relevant in the limit of large kinetic energies, quite the opposite of the ultracold collision regime. However, the RRKM estimate for the lifetime is the same at any energy where $N_o=1$. We can thus compare classical trajectory lifetimes with the RRKM estimate without performing classical trajectories in the Wigner threshold regime. In addition, in the present problem we are concerned with the motion of the atoms deep inside the potentials where their kinetic energy is, in fact, large.  They spend comparatively little time getting into and out of the collision complex. The classical lifetimes are therefore expected to represent the appropriate time delay one would find by propagating wave packets.

\subsection{Potential Energy Surface}
In this work all calculations were performed on the quartet surface, assuming both the atom and the molecule are spin-polarized and that spin plays no role in the dynamics of the complex. For this calculation we use a pairwise-additive three-atom potential based on Lennard-Jones (LJ) atom-atom pair potentials
\begin{equation}
\label{eqn:PES}
V(\bm{r_1},\bm{r_2},\bm{r_3}) = \sum_{i \ne j} V_{LJ}(\bm{r_i}-\bm{r_j}),
\end{equation} 
where
\begin{equation}
\label{eqn:LJ}
V_{LJ}(r) = \frac{C_{12}}{r^{12}} - \frac{C_6}{r^{6}}.
\end{equation}
We use a realistic $C_6$ for the atom-atom pair potentials and choose the $C_{12}$ such that the LJ potential has the correct atom-atom depth, $D_e$ ($C_{12} = C_6^2/4D_e$). To span a range of masses and interactions, we construct surfaces for three systems of current experimental interest, $^7$Li, $^{39}$K and $^{133}$Cs, whose values for the $C_6$ and $D_e$ are shown in table \ref{tab:C6:De}. This simple choice of potential ignores 3-body terms however is sufficient for the exploratory nature of this work.
\begin {table}[b]
\begin{center}
\begin{tabular}{| c || c | c |} 
  \hline 
  System & $C_6$ (a.u.) & $D_e$ (cm$^{-1}$) \\
  \hline  
  \hline                       
  Li+Li & 1394\cite{Dattani:2011}  & 334\cite{Dattani:2011}  \\
  K+K   & 3927\cite{Regal:2003}    & 253 \cite{Zhao:1996}     \\
  Cs+Cs & 6891\cite{Xie:2009}     & 279 \cite{Xie:2009}      \\
  \hline
\end{tabular}
\caption {\label{tab:C6:De}Van der Waals coefficients $C_6$ and well depths $D_e$ for the triplet states of Li$_2$, K$_2$ and Cs$_2$. }
\end{center}
\end {table}

\subsection{Lifetimes}

Each classical trajectory leads to a different lifetime, defined in our calculations as follows.  For a given initial condition the lifetime was computed as the time difference between the collision complex forming and breaking up. The collision complex was considered formed when the hyper radius $\sqrt{R_{AB}^2+R_{BC}^2+R_{AC}^2}$ is first less than $\sqrt3\bar a$ where $R_{AB}$ is the distance between atoms A and B. $\bar a$ is the characteristic length scale of the potential as defined by Gribakin and Flambaum for a potential varying as $-C_n/R^n$ \cite{Gribakin:1993}. The collision complex was considered to have broken up when the hyper radius was again bigger than $\sqrt 3 \bar a$ and the collision partners have enough kinetic energy to escape to $R_{\infty}$. In this way the dwell time is associated with the short-range physics dominated by fast semiclassical motion, and shorn from the details of long-range motion that are best handled quantum mechanically at ultralow  collision energies.

The time for a lone atom to cover a distance $\bar a$ in the absence of a potential is 0.05, 0.94 and 7.96 ns for Li + Li$_2$, K + K$_2$ and Cs + Cs$_2$ at an energy corresponding to the lowest rotational threshold of each system. In this work the lifetime is dominated by complex short-range behaviour, as such the explicit lifetime as computed differs negligibly from the time delay defined as the difference between the dwell times of a classical trajectory computed with and without the interaction potential \cite{Smith:TimeDelay:1960}.

\section{Results and Discussion}
\subsection{Density of States and Lifetimes}
The primary outcome of the statistical model proposed by Mayle {\it et al.} is the long dwell time of the complex. Within that theory a lifetime estimate is unambiguously assigned a single number for a given density of states. We compute this lifetime using equation \ref{eqn:rrkm} and estimating the DOS, $\rho$ as explained in detail in \cite{Mayle:2012,Mayle:2013}. The single channel  Schr\"{o}dinger equation was solved using the Fourier grid Hamiltonian method \cite{Balint-Kurti:FGH:1992,Marston:FGH:1989} using the same LJ potential as for the classical trajectories. As with the PES for the classical trajectories the potential is assumed to be pairwise additive with $C_6$ and $D_e$ chosen to be double the atom + atom value for the atom + dimer potential. The estimated DOS shown here does not include the factor of 6 reduction due to identical bosons in order to allow direct comparison with the classical trajectory estimate for the lifetime.

\begin{figure}[tbp]
\centering
\includegraphics[width=1.0\columnwidth]{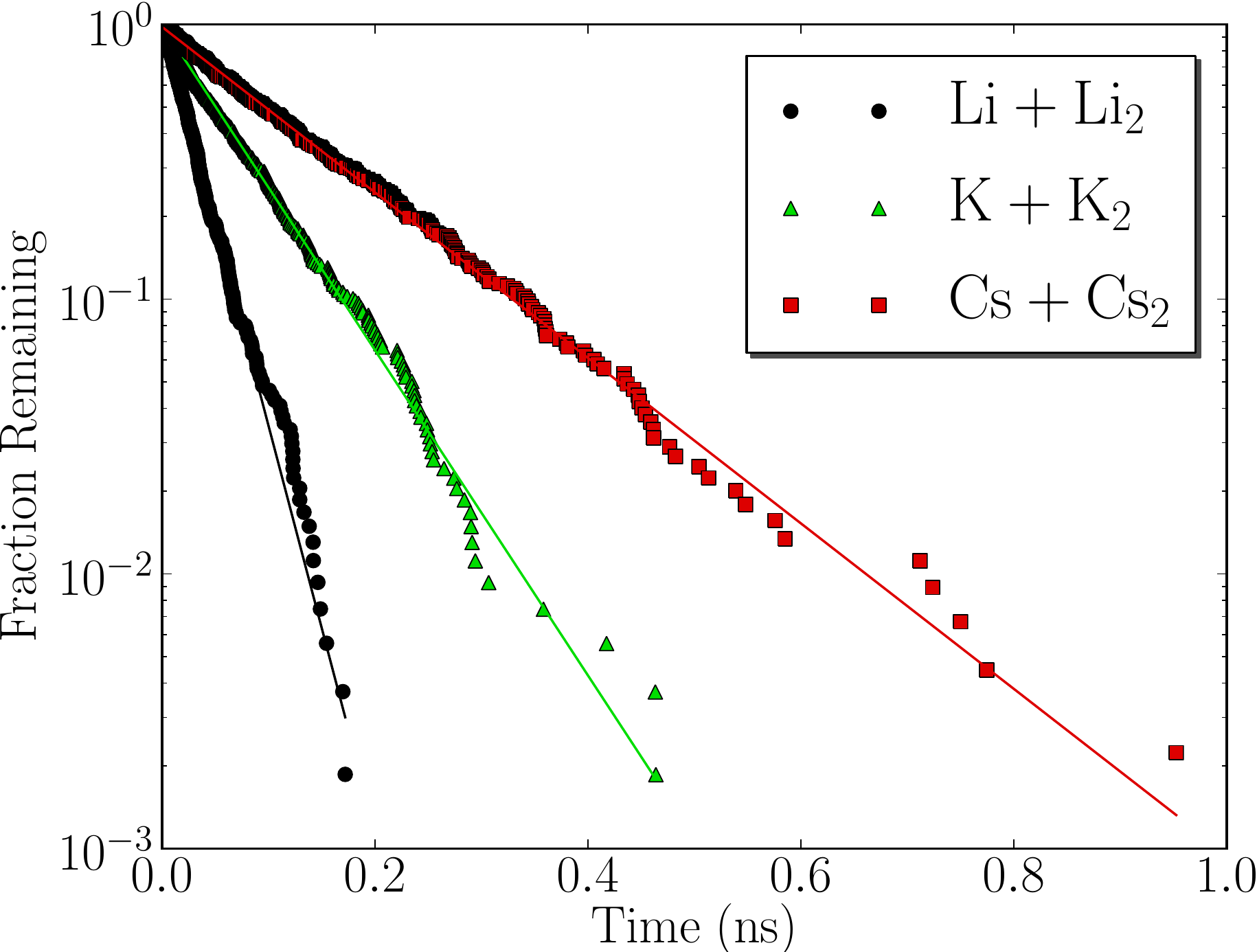}
\caption{(Color online) The fraction of collision complexes yet to decay as a function of time at a collision energy of 30~K for Li + Li$_2$, K + K$_2$ and Cs + Cs$_2$. The solid lines show an exponential decay fitted to the data points for each system. For each system over 1000 trajectories were run for random initial $\theta$ between 0 and $\pi / 2$ with no impact parameter.} 
\label{fig:tau_exp_decay}
\end{figure}

In the present classical calculations, each trajectory has its own dwell time, and these vary wildly with initial condition. Nevertheless, if the RRKM assumption of ergodicity of the trajectories holds, it follows that the lifetimes are distributed according to an exponential distribution
\begin{equation}
\label{eqn:lt_exp_decay}
f = \exp\left(-{\frac{t}{\bar\tau_{ct}}}\right),
\end{equation} 
where $f$ is the fraction remaining after time $t$ and $\bar\tau_{ct}$ is the average lifetime. We obtain the mean lifetime for a given collision energy by running a large number of trajectories and computing the number remaining within the complex, as a function of time. Results for this fraction are shown in Figure \ref{fig:tau_exp_decay} for all 3 systems, showing data from more than 1000 trajectories calculations for each. To a good approximation, the fraction is an exponential function of time, justifying the approximation in equation (\ref{eqn:lt_exp_decay}).

The exponential decay of this fraction can itself be understood using statistical arguments. Any particular trajectory at low collision energy that remains within the collision complex can be interpreted as consisting of a large number of individual mini-collisions, each of which essentially randomizes the energy distribution among the three atoms.  A very small fraction of these mini-collisions results in fragmentation.  In an ensemble of trajectories with differing initial conditions, the number of trajectories able to escape the complex at any given time is therefore proportional to the number that have not yet escaped by this time.  This proportionality leads to the exponential dependence.  We interpret this dependence as evidence that the collision complex, viewed classically, explores large, regions of phase space randomly, as asserted by the statistical theory. This criterion, of exponential lifetime distribution, is interpreted within the RRKM theory as a signature of trajectories that fill phase space ergodically \cite{Hase:2005}.  Here we adopt this interpretation as evidence for ergodicity, although we have not attempted to calculate the filling of phase space directly. At higher collision energies where there is of order one collision event within the complex, the use of Equation \ref{eqn:lt_exp_decay} is no longer valid. In the systems studied here this corresponds to collision energies above about 400 K.

\begin{figure}[tbp]
\centering
\includegraphics[width=1.0\columnwidth]{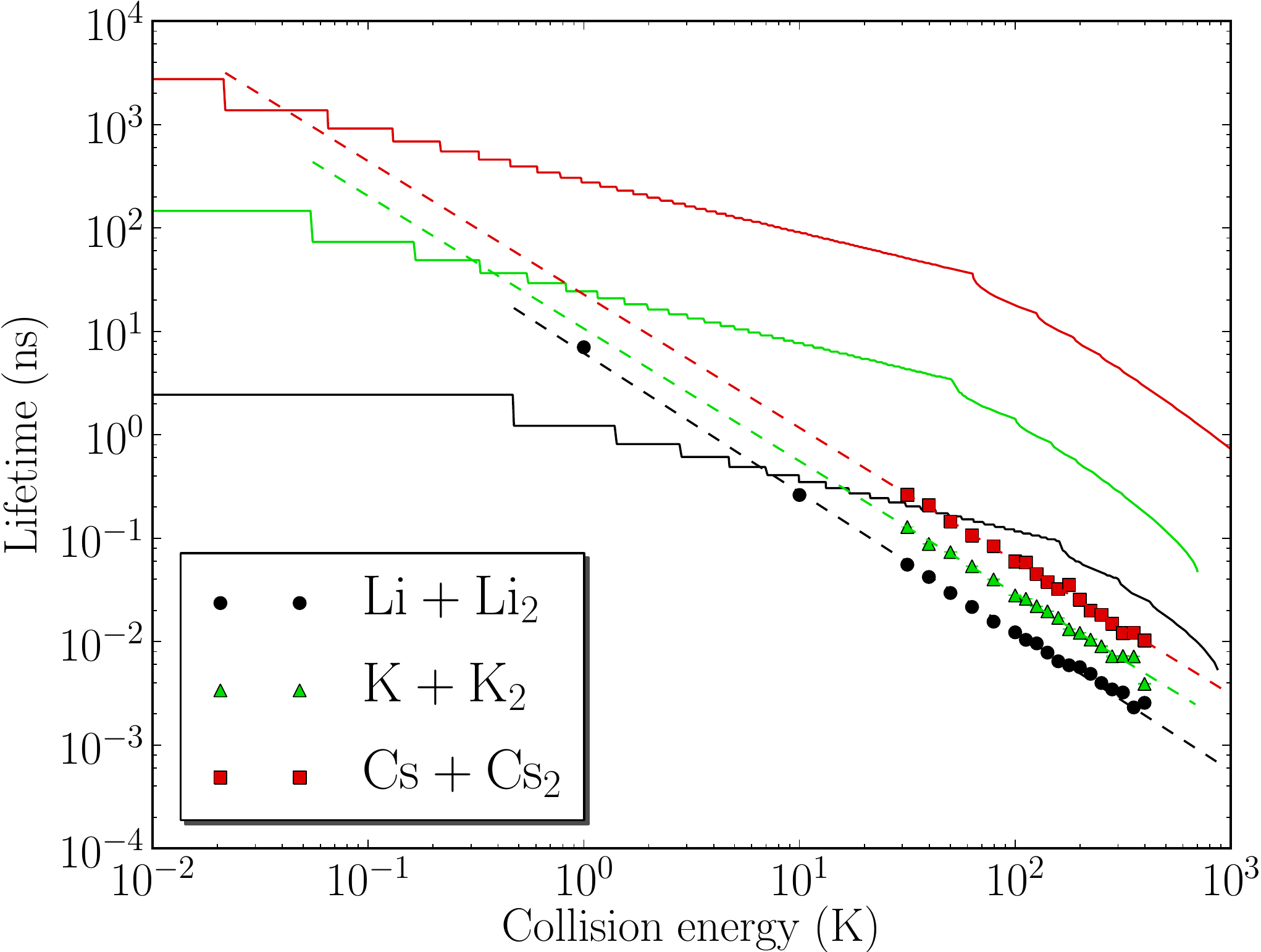}
\caption{(Color online) Lifetime as a function of collision energy for the collisions of Li + Li$_2$, K + K$_2$ and Cs + Cs$_2$. Shown are lifetimes computed from explicit dwell times from classical trajectories (dots) and from the RRKM approximation (solid line). The dotted line shows a power law fit to the classical trajectory data.} 
\label{fig:lt_ALL}
\end{figure}

Armed with a clear definition of the initial-state-ensemble averaged lifetime, we now ask what is the energy dependence of this lifetime,  in particular in the ultracold limit.  To this end, Figure \ref{fig:lt_ALL} shows how the lifetime scales with collision energy for collisions of all three species. In this figure the solid line is computed using the RRKM formula (\ref{eqn:rrkm}), with density-of-states computed according to the algorithm of Mayle.  For comparison, the points connected by dotted lines are the lifetimes as computed from classical trajectories by the methods just outlined.

The lifetime computed from classical trajectories shows a power-law dependence with energy over the range shown (with exponent -1.35, -1.28 and -1.29 for Li + Li$_2$, K + K$_2$ and Cs + Cs$_2$ respectively) and will extrapolate to infinite lifetime in the zero-energy limit. This is appropriate for the classical calculation, since the classical phase space into which the complex can decay shrinks without limit as $E_{col} \rightarrow 0$. In a realistic quantum system, by contrast, the phase space corresponding to the fragmentation can only meaningfully diminish until a single channel remains (neglecting Wigner threshold law effects).  For this reason, the most viable comparison between the calculations is the lowest rotational threshold for the molecule, in this case at collision energy  0.62, 0.04 and 0.005 K for Li + Li$_2$, K + K$_2$ and Cs + Cs$_2$ respectively.

The power law dependence of the lifetime $\bar\tau_{ct}$ as a function of collision energy in Figure \ref{fig:lt_ALL} affords extrapolation of this lifetime to low collision energies.  This is a useful procedure when the lifetimes become so large as to be computationally burdensome.  We use this extrapolation to arrive at lifetime estimates for the larger molecules. This estimate for the lifetime at the energy of the first excited ro-vibrational level compared to the DOS estimate for our 3 different systems is shown in table \ref{tab:dos:tau}. It is seen that the estimates are all in agreement to better than an order of magnitude. The agreement is better for heavier systems which have smaller rotation splitting leading to much longer lifetimes. This good agreement strongly indicates that the lifetimes estimated in \cite{Mayle:2013}, for collisions for alkali-molecule + alkali-molecule systems, are also realistic. We have not, however, performed four-atom classical scattering calculations to test this idea.  Lifetime calculations for collisions including a nonvanishing randomly chosen impact parameter $b$ were performed and found to lead to lifetimes consistent with collisions with $b=0$.
\begin {table}[b]
\begin{center}
\begin{tabular}{| c || c | c | c |} 
  \hline 
  System & DOS(mK$^{-1}$) & $\tau_{dos}$(ns) & $\bar{\tau}_{ct}$(ns)  \\
  \hline  
  \hline                       
  Li+Li$_2$ & 0.05  & 2.4  & 12    $\pm$ 1  \\
  K+K$_2$   & 3.04  & 146  & 303  $\pm$ 36  \\
  Cs+Cs$_2$ & 57.22 & 2746 & 2871 $\pm$ 328 \\
  \hline
\end{tabular}
\caption{\label{tab:dos:tau} Ro-vibrational DOS(mK$^{-1}$) and lifetimes at ultralow collision energies of collision complexes $\tau$(s) for Li + Li$_2$, K + K$_2$ and Cs + Cs$_2$ from both the DOS method and Classical trajectory calculations.}
\end{center}
\end {table}

The lifetimes obtained are self consistent, as such we would expect that the lifetime obtained with a more realistic potential would not change our predictions much as the DOS is not sensitive to details of the potential. This further emphasizes the utility of the RRKM estimate for the lifetime, since the DOS is not sensitive to details of the potential an estimate can be made for the lifetime without needing a full accurate potential for each system of interest. In this work only the depth, $D_e$, and $C_6$ for the atom-atom potential were needed to obtain the lifetime estimate for each system. Since these are known or can be estimated for all the alkali pairs it is relatively simple to provide an order of magnitude estimate for the lifetime of a given system of interest.

While the two approaches agree closely at energies where there are only a couple of open channels, at higher collision energies the RRKM formula tends to overestimate the lifetime as compared to the classical calculation. This is because the RRKM estimate of the lifetime assumes that the collisions are ergodic, so that the $\rho$ in estimate (\ref{eqn:rrkm}) is the density of all states that satisfy angular momentum conservation.  However at higher energies, there are so many exit channels that the complex decays before exploring all of the available phase space, reducing the value of the effective DOS $\rho$. Turning this around we can interpret the agreement of the lifetimes at low energies as evidence that the dynamics is ergodic in this limit.

\begin{figure}[tbp]
\centering
\includegraphics[width=0.65\columnwidth]{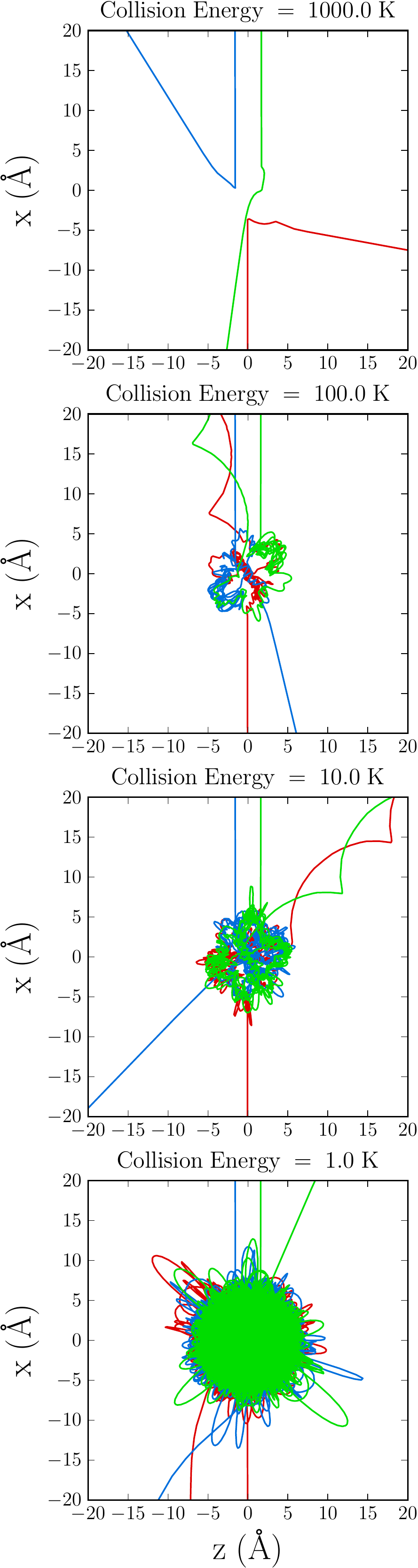}
\caption{(Color online) Representative trajectories for collisions of Li + Li$_2$ over a range of collision energies. The different colors labeling different atoms. All calculations were performed without impact parameter for $\theta/\pi = 0.25$.} 
\label{fig:trajectory}
\end{figure}

To better illustrate the complexity of the trajectories, figure \ref{fig:trajectory} shows representative trajectories for Li + Li$_2$ over a range of collision energies. This figure shows the increasing length of trajectories as the collision energy decreases, with lower panels showing lower collision energies.  While these sample trajectories do not prove that low-energy  scattering is ergodic, they do certainly show that high-energy scattering is {\it not} ergodic.  Rather, at higher collision energies, only a handful of mini-collisions occur before the collision fragments separate.

Thus at energies well above threshold, the RRKM expression may be expected to overestimate lifetimes, as seen in Figure  \ref{fig:lt_ALL}.  The trend of lifetime versus collision energy is also comprehensible.  At energies below the first vibrational threshold of the molecule, the number of open channels increases according to the rotational spacing $\propto BN(N+1)$, where $B$ is the rotational constant of the molecule, and $N$ its rotational quantum number.  Thus the number of open channels should scale as $N_o \propto E_{col}^{-1/2}$, leading to $\tau_{dos} \propto E_{col}^{-1/2}$, which is indeed the scaling of the RRKM lifetime in Figure \ref{fig:lt_ALL} in this energy regime.  By contrast, in this regime, the increase in classical phase space into which the complex  decays  includes both vibrational and rotational degrees of freedom, since vibration need not be quantized.  This  leads to a faster apparent growth in phase space classically, and a faster decay of classical lifetime as compared to RRKM.

At collision energies well above the first vibrational excitation of the molecule, the RRKM  and classical lifetimes diminish at more closely matched rates.  In this regime, where many more exit channels, both rotational and vibrational, are open the counting argument for $N_o$ seems to accord better with the opening up of phase space as $E_{col}$ grows.   This is most clearly seen in the two heavier species.  Still, as noted above, the absolute lifetime is overestimated by the RRKM expression, since the collisions clearly do not explore the full phase space implied by $\rho$.

\subsection{Onset of Chaos}
Given the complexity of long-time trajectories at low collision energy, one suspects that classical chaos is at work. Chaos is of fundamental interest, unifying a wide array of disparate topics from the motion of planets, turbulent fluid flow through to the predication of the weather and the economy. Inherently nonlinear phenomena such as these can appear to be intractably complicated, however when viewed through the lens of chaos exhibit an orderliness which provides deep and unifying insight. Classically chaotic systems leave signatures in the corresponding quantum-mechanical system via the Gutzweiller trace \cite{Gutzwiller:1971}. Chaos in quantum systems manifests itself statistically in a number of  ways such as the Wigner-Dyson distribution of energy level intervals \cite{Wigner:1951,Wigner:1955,Wigner:1957,Wigner:1958,Dyson:1:1962,Dyson:2:1962,Dyson:3:1962}, Porter-Thomas statistics of resonance widths \cite{Porter:1956} and Ericson fluctuations \cite{Ericson:1960,Ericson:1963}.

Classical chaotic scattering is a manifestation of transient chaos where particles move freely before and after collision events however during the collision event the particles are strongly interacting and the motion can be chaotic. Such collisions have been extensively studied in the context of chemical reaction dynamics \cite{Rankin:1971,Kovacs:1995,Barr:2009} and cold collisions \cite{Atkins:1995,Pattard:1998,Bohn:fesh:2002}. The route to chaos in classical scattering has also been studied in a variety of different scattering systems, where chaotic effects are seen to arise suddenly below a critical energy \cite{Bleher:1990,Pattard:1998,Schelin:2008}.

\begin{figure}[tbp]
\centering
\includegraphics[width=1.0\columnwidth]{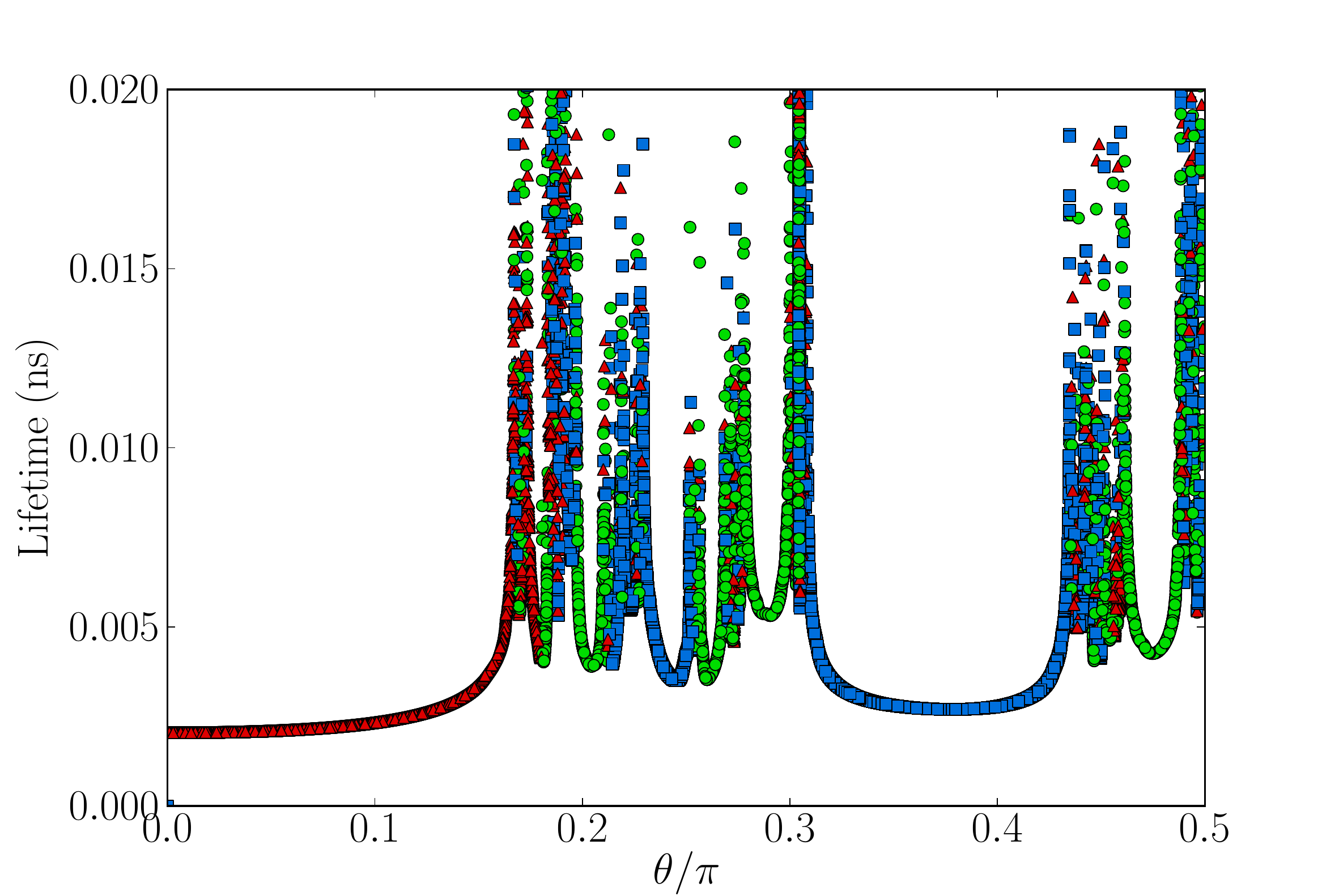}
\includegraphics[width=1.0\columnwidth]{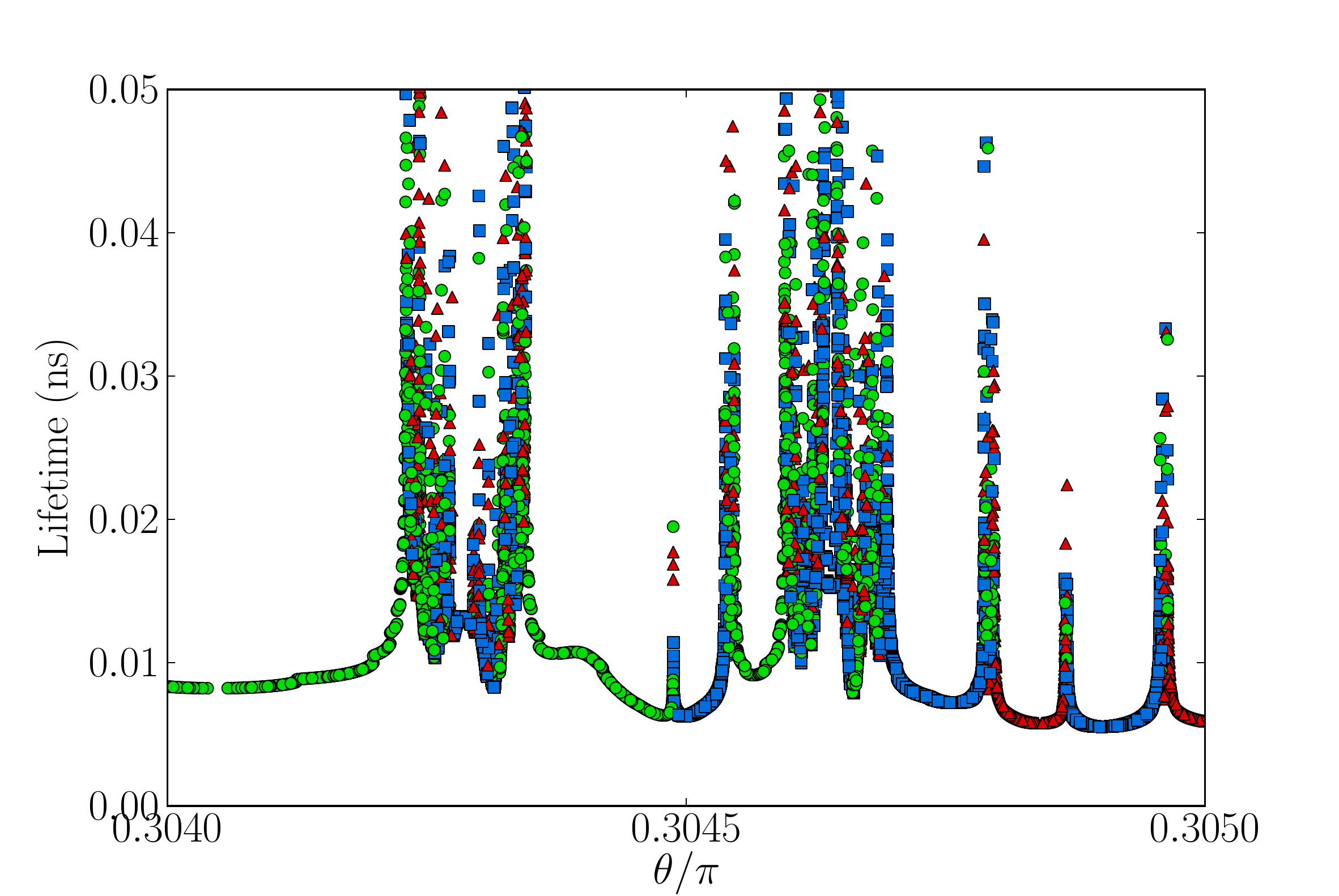}
\caption{(Color online) Lifetime as a function of initial $\theta$ for collisions of Li + Li$_2$ with zero impact parameter at $E_{col} = 450$ K. Different colors correspond to different final basins for the trajectory. The lower panel shows a 500 times magnified region of the upper panel.} 
\label{fig:self_similar}
\end{figure}

To illustrate the presence of chaos in our classical simulations, we show in Figure \ref{fig:self_similar} the single-trajectory lifetime of the collision complex for Li + Li$_2$ as a function of the initial angle $\theta$, with impact parameter $b=0$. The three colors label trajectories which finish in different final ``basins,'' corresponding to which of the three atoms emerges freely after fragmentation of the complex. (Recall that in the classical simulation the atoms are regarded as distinguishable). Some regions of initial $\theta$ lead to collisions with similar short lifetimes, and to the ejection of a particular atom. These are regions where there is a single mini-collision event, after which one of the atoms has enough kinetic energy to escape. Collisions in these  regions all follow a similar trajectory reflected in the same pair of atoms composing the dimer at the end. In other regions the lifetime is longer and varies rapidly as a function of initial $\theta$. These are regions where there are multiple mini-collision events, in which the energy is redistributed until one of the atoms has attained enough kinetic energy to escape.

The lower panel of Figure \ref{fig:self_similar} is a 500 times magnified region of the upper panel. It is qualitatively similar to the upper panel, despite the vast disparity in scale of angle shown. This scale invariance is a feature of fractals and is characteristic of chaotic scattering \cite{Bleher:1990,Ott:1993}. Qualitatively this scale invariance implies a set of singularities in figure \ref{fig:self_similar}, which are well understood and correspond physically to initial conditions which enter the scattering region and never leave \cite{Noid:1986,Atkins:1995}.

The set of singularities implied by the scale invariance exhibited in Figure \ref{fig:self_similar} can be quantified by a fractal dimension \cite{Mandlebrot:1967} using a procedure charmingly named the uncertainty algorithm \cite{McDonald:1985}. In this algorithm, trajectories are classified as stable under perturbation $\delta$ if two trajectories differing in initial condition by $\delta$ finish in the same basin. In this work final basins correspond to the three possible collision outcomes AB + C, AC + B or BC + A. If this is not the case then the trajectory is considered unstable under perturbation $\delta$. Running a large number of random initial conditions differing by $\delta$ the fraction of unstable initial conditions for a given $\delta$, denoted $f(\delta)$, can be computed.

\begin{figure}[tb]
\centering
\includegraphics[width=1.0\columnwidth]{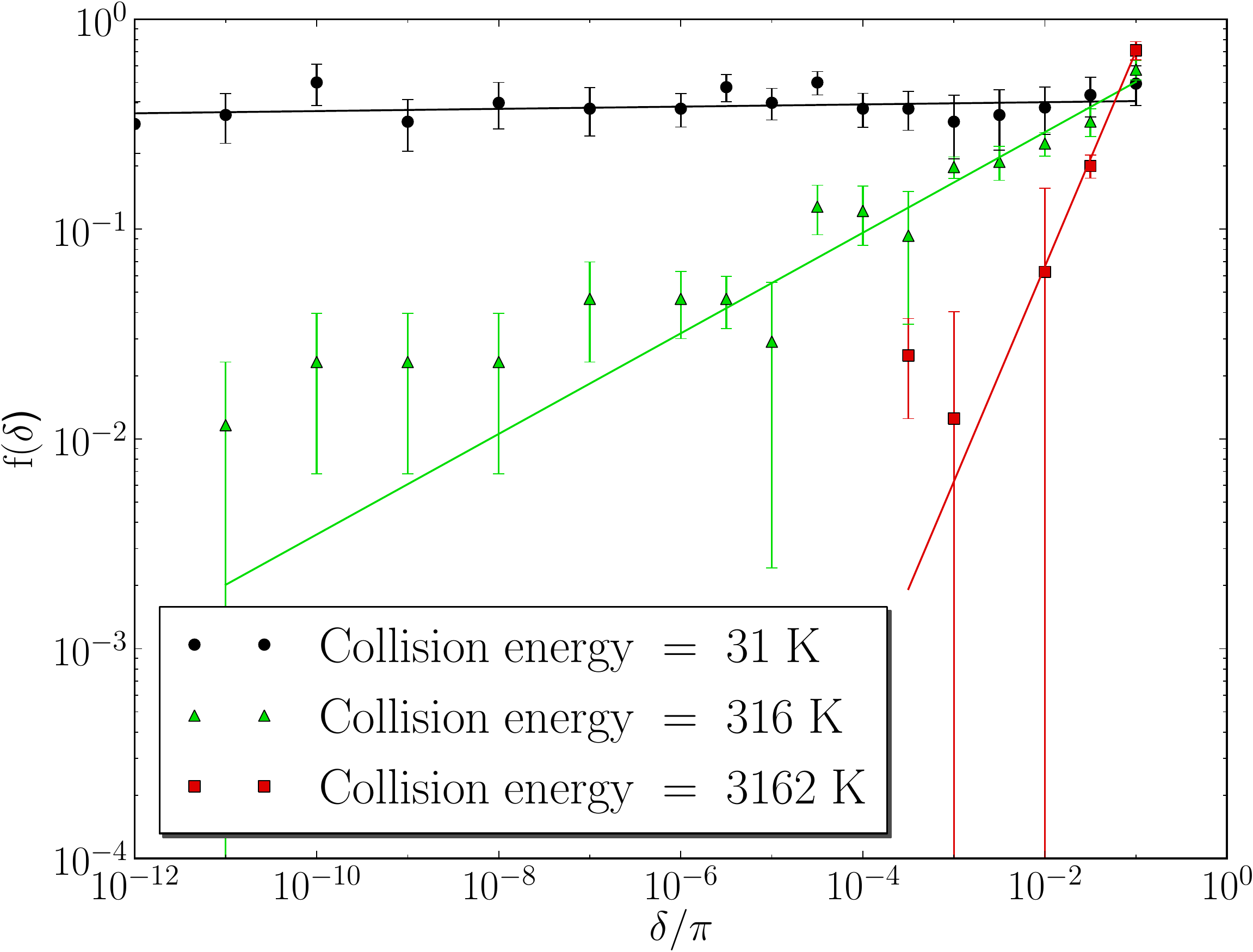}
\caption{(Color online) $f(\delta)$ as a function of $\delta$ for collision of Li + Li$_2$. Shown are three representative collision energies demonstrating the full range of behavior. The corresponding $\alpha$ values are 0.01 at 31 K, 0.24 at 316 K and 1.00 at 3162 K.} 
\label{fig:fraction_delta}
\end{figure}

Figure \ref{fig:fraction_delta} shows this fraction as a function of $\delta$ at three different collision energies for collisions with $b=0$. The behavior seen in \ref{fig:fraction_delta} is characterized by the uncertainty algorithm,
\begin{equation}
 \label{eqn:un_exp}
 f(\delta,E_{col}) \propto \delta^{\alpha(E_{col})}.
\end{equation} 
where $\alpha$ is the uncertainty exponent. At high collision energies, the unstable fraction decreases rapidly as a function of $\delta$, quantified in equation (\ref{eqn:un_exp}) by $\alpha = 1$. This is because there are many regions of initial conditions $\theta$ where all trajectories within $\delta$ of $\theta$ finish in the same basin. As the collision energy is lowered, however, even small steps in $\delta$ can lead to completely different final basins for many initial values $\theta$, quantified by $\alpha$ decreasing from 1. At the very lowest energies shown,  the unstable fraction no longer depends of $\delta$ at all at which point $\alpha = 0$. At such low collision energies the outcome of two collisions whose initial conditions differ by an arbitrarily small amount are unrelated, like the toss of a (three-sided) coin. This unpredictability again suggests that during a collision event the total energy is redistributed randomly between the degrees of freedom of the system.  At lower collision energies there is less energy to go round and so the probability of a single atom having enough after each collision event to escape is lower. Thus the fraction of trajectories which are unstable under perturbation $\delta$ is higher at lower collision energies where neighboring trajectories have longer to diverge.

The exponent $\alpha$ can be given a geometrical interpretation based on basin boundaries. We can divide up regions of initial conditions in $\theta$ by which final basin they end in. Such regions can be seen in Figure \ref{fig:self_similar} as regions of a single color. The fractal dimension $d$ of the boundary between such regions is related to the uncertainty exponent $\alpha$ by
\begin{equation}
 \alpha = D - d,
\end{equation}
where $D$ is the dimension of initial phase space associated with perturbation $\delta$, in this case where $\delta$ explores the single degree of freedom  $\theta$, $D=1$ \cite{McDonald:1985,Ott:Chaos:2002}. $\alpha$ can thus take values between 0 and 1 since the dimension of the boundary basin can be at most 1 less than the dimension of phase space. Thus as $\alpha$ decreases the fractal dimension of the boundary between different final basins increases. As this happens regions leading to the same final basins shrink and small differences in initial conditions can put neighboring trajectories in different final basins regardless of initial condition. Eventually when $\alpha = 0$ the basin boundary fills the entire space. When this happens all initial conditions lie on a basin boundary leading to completely different trajectories from their neighbors, on any arbitrary length scale.
\begin{figure}[bt]
\centering
\includegraphics[width=1.0\columnwidth]{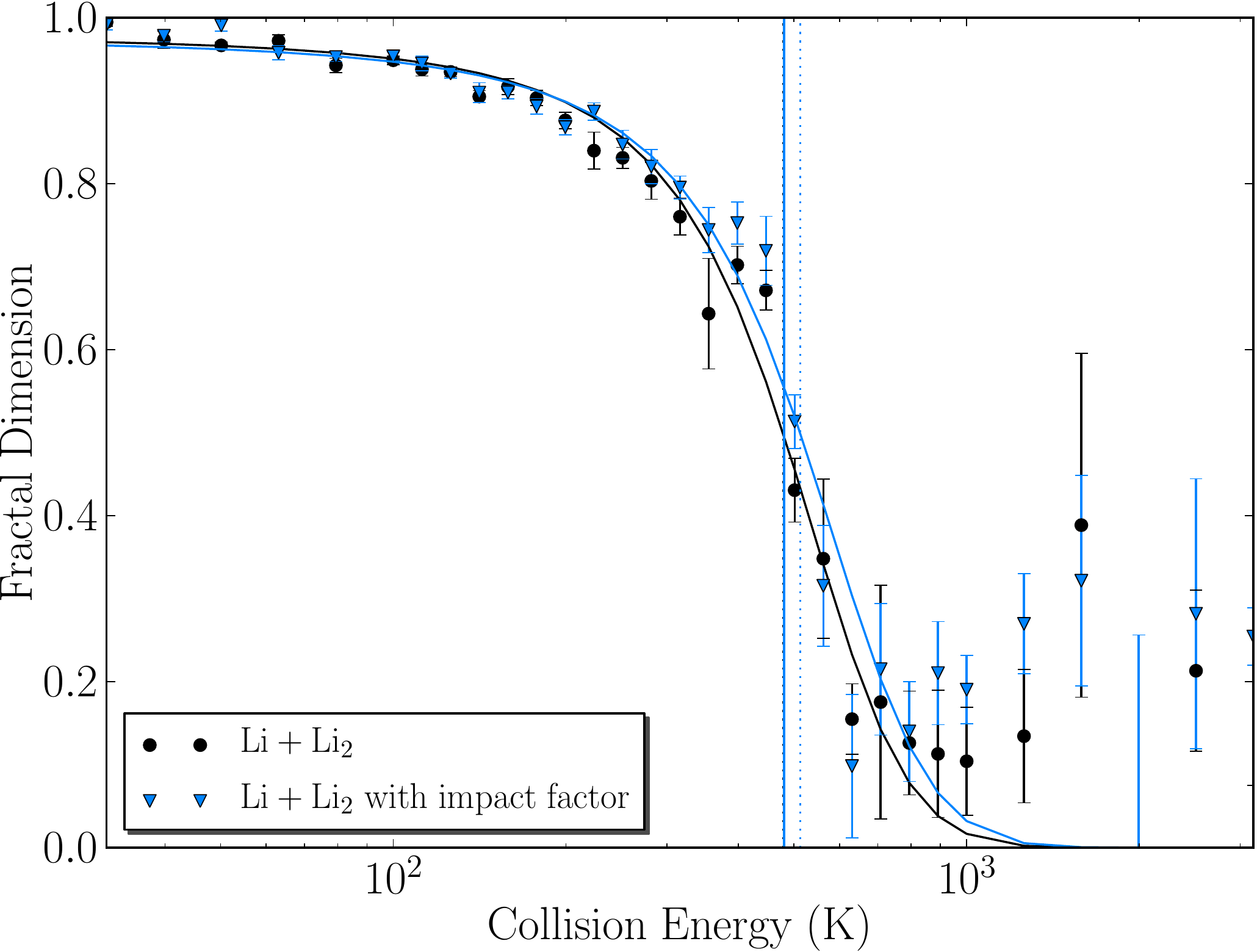}
\caption{(Color online) Fractal dimension, $d$, as a function of collision energy for collisions of Li + Li$_2$ both with and without an impact parameter. $d-2$ is plotted for the result including an impact parameter for comparison. The vertical dotted line is $E_{chaos}$ as defined in equation \ref{eqn:onset} and the vertical solid line is the atom-atom well depth.}
\label{fig:fractal_dimension_Li_Li_b}
\end{figure}

Figure \ref{fig:fractal_dimension_Li_Li_b} shows the fractal dimension $d$ as a function of collision energy for Li + Li$_2$ both with ($D=3$) and without ($D=1$) an impact parameter. The impact parameter was uniformly randomly chosen such that the total angular momentum was between 0 and $\hbar/2$ (s-wave collisions). The solid lines in the figure are fits to a switching function of the form
\begin{equation}
\label{eqn:onset}
 d=\frac{1}{2}\tanh\left(\frac{E_{chaos}-E_{col}}{\Gamma_c}\right) + D -\frac{1}{2}
\end{equation}
where $E_{chaos}$, the point of inflection, gives the energy of the onset of chaos and $\Gamma_{chaos}$ defines a width. For Li with nonzero or zero impact parameter these are 478 $\pm$ 16~K and 512 $\pm$ 24~K respectively where the error given is 1 standard deviation. It is seen that for collisions of Li with Li$_2$ both with and without an impact parameter these values are the same to within 1 standard deviation. We thus conclude that the onset of chaos is independent of the impact parameter for a given collision system.  Therefore, in computing $d$ for heavier species, we are justified in setting $b=0$, which simplifies the calculations.

\begin {table}[b]
\begin{center}
\begin{tabular}{| c || c | c | c |} 
  \hline 
  System & $D_e$~(K) & $E_{chaos}$~(K) & $\Gamma_{chaos}$~(K) \\
  \hline  
  \hline                       
  Li+Li$_2$ & 480  & 478 $\pm$ 16 & 256 $\pm$ 13 \\
  K+K$_2$   & 364  & 371 $\pm$ 11  & 209 $\pm$ 11 \\
  Cs+Cs$_2$ & 402  & 401 $\pm$ 17  & 220 $\pm$ 15 \\
  \hline
\end{tabular}
\caption {\label{tab:a:b:de} The atom-atom well depth onset of chaos and width of transition for Li + Li$_2$, K + K$_2$ and Cs + Cs$_2$. The error the standard deviation of the parameter estimate from the least squares fitting.}
\end{center}
\end {table}

\begin{figure}[tb]
\centering
\includegraphics[width=1.0\columnwidth]{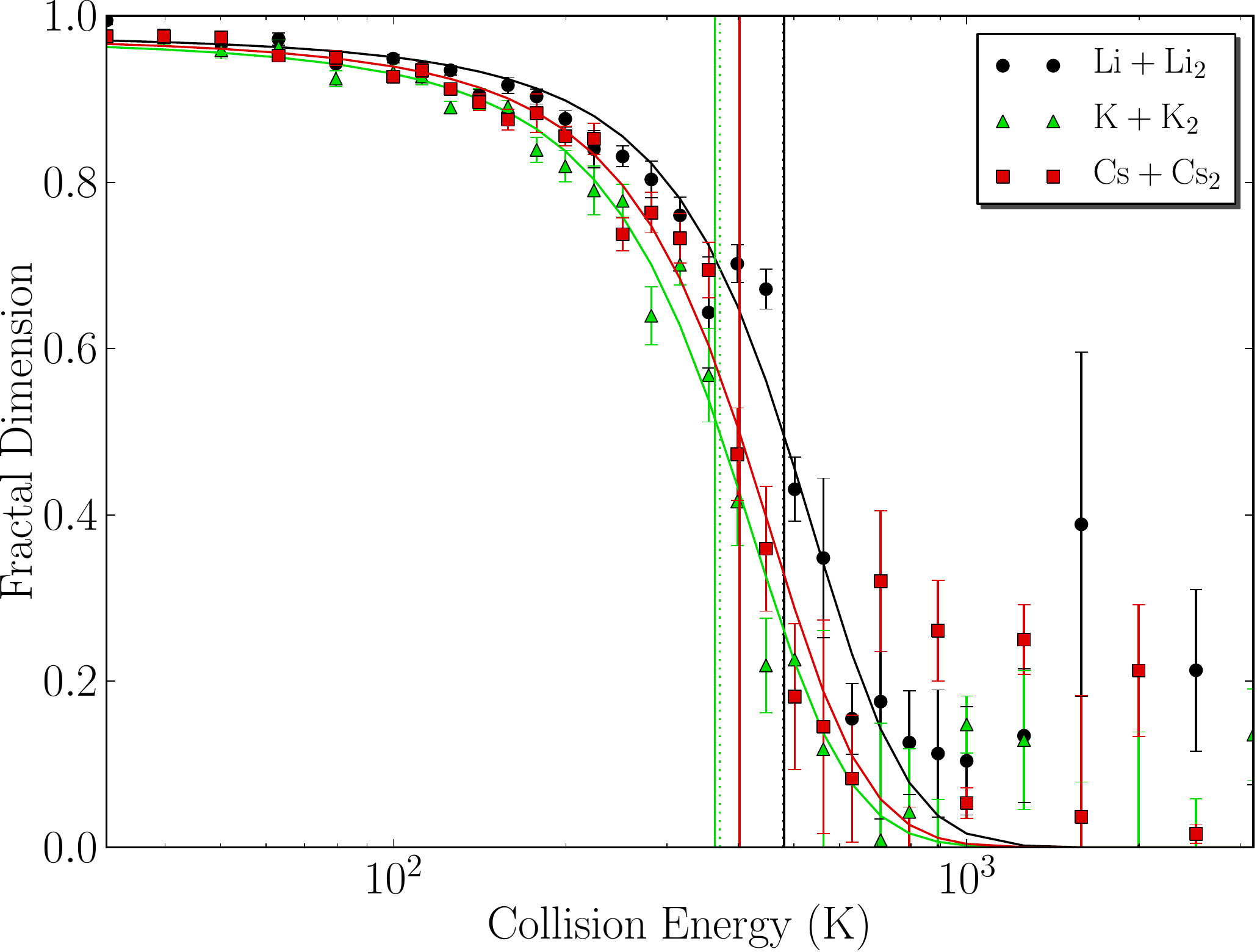}
\caption{(Color online) Fractal dimension, $d$, as a function of collision energy for collisions of Li + Li$_2$, K + K$_2$ and Cs + Cs$_2$ for collisions without an impact parameter. The vertical dotted line is $E_{chaos}$ as defined in equation \ref{eqn:onset} and the vertical solid line is the atom-atom well depth for each species.} 
\label{fig:fractal_dimension_Li_Cs_K}
\end{figure}

Figure \ref{fig:fractal_dimension_Li_Cs_K} compares the fractal dimension, $d$, as a function of collision energy for our three systems Li + Li$_2$, K + K$_2$ and Cs + Cs$_2$ without an impact parameter. It is seen that collisions become chaotic as the collision energy becomes lower, there is a sudden increases in the fractal dimension when the collision energy becomes less than the atom-atom well depth (shown as vertical solid lines). Values for $E_{chaos}$ and $D_e$ for the three systems are shown in table \ref{tab:a:b:de}. At collision energies below the atom-atom well depth the dimer is able to absorb enough of the lone atom kinetic energy into its internal degrees of freedom to prevent it escaping. Above this energy the lone atom is able to dissociate the dimer and still has energy left over. We thus conclude that molecular collisions at sub-microkelvin temperatures achieved experimentally are chaotic. This justifies the assumption made by Mayle {\it et al} that resonances, if resolved, should obey nearest-neighbor statistics associated with quantum chaos, such as the Gaussian orthogonal ensemble (GOE). This further justifies the use of Equation \ref{eqn:lt_exp_decay} to compute the mean lifetime as the time delay statistics for chaotic scattering decay exponentially. Such an exponential decay is characteristic of hyperbolic scattering where all periodic orbits are unstable \cite{Ott:1993,Ott:Chaos:2002}. The lack of stable periodic orbits in the system is a necessary condition for a system to be ergodic as stable orbits only explore their own region of phase space. With no stable periodic orbits the system is ergodic in the limit $E_{col} \rightarrow 0$ where $\bar\tau_{ct} \rightarrow \infty$, further supporting our conclusion that collisions at sufficiently low energy, achievable experimentally, are ergodic.

In this work we have used a simple pairwise additive model for the quartet potential, however the alkali metal trimer are highly nonadditive \cite{Soldan:2003}. Despite this we would not expect that using a more realistic potential would change this prediction as the onset of chaos is primarily determined by the atom-atom well depth, correct in our model, and not details of the potential surface. We would also predict that for collisions on the doublet surface where the dimer is in a singlet state the onset of chaos would occur at about the singlet well depth which is generally much deeper than the triplet depth. It should be noted that the predictions for the onset of chaos made here are all at collision energies many orders of magnitude higher than the sub-microkelvin temperature achieved experimentally.

\subsection{Relevance to lifetime calculations}

The dominant role of classical chaos at low collision energy also has implications for the applicability of RRKM ideas.  The RRKM lifetime (\ref{eqn:rrkm}) would overestimate lifetimes if somehow not all of the phase space $\rho$ were accessed in collisions \cite{Hase:2005}.  However, classical chaos as a function of initial condition implies that, averaged over initial conditions, the trajectories access wildly different regions of phase space, so that all of $\rho$ is likely to contribute. This surmise is consistent with the lifetime agreement in Fig. \ref{fig:lt_ALL}, where many trajectories with varied initial conditions are calculated.

\section{Conclusions}
In the present work we have performed classical trajectory calculations that yield explicit dwell times consistent with the simple RRKM estimates at low collision energies for three systems of current experimental interest, $^7$Li + $^7$Li$_2$, $^{39}$K + $^{39}$K$_2$ and $^{133}$Cs + $^{133}$Cs$_2$. Lifetimes were compared for collisions on an approximate quartet surface, assuming both the atom and the molecule are spin-polarized and that spin plays no role in the dynamics of the complex. The agreement of these results is extremely promising as it indicates that lifetime estimates for alkali atom-dimer collisions on the doublet surface and alkali dimer-dimer collisions are also well approximated by the simple RRKM estimate for the lifetime. Such predictions have already been made where the lifetime of the complex was found to be long, of order 10-100 ns for alkali atom-alkali dimer collisions, and of order 1-10 msec for collisions of alkali molecules with one another. Such long lifetimes are comparable to experimental lifetimes, and may lead to novel trap loss mechanisms \cite{Mayle:2012,Mayle:2013}. We interpret the agreement of the lifetimes at low energies, as well as their exponential distribution, as evidence that such collisions are ergodic.

Further we found that low energy collisions exhibit chaos at collision energies lower than the atom-atom binding energy. We quantify the onset of chaos in terms of a ``fractal dimension'' for the space of incident conditions, finding that classical chaos emerges well above ultracold energies. This justifies applying chaotic arguments when studying ultracold collisions \cite{Flambaum:2006,Mayle:2012,Mayle:2013}. Classically chaotic systems leave signatures in the corresponding quantum-mechanical system via the Gutzweiller trace \cite{Gutzwiller:1971}. Chaos in quantum systems manifests itself statistically in a number of  ways such as the Wigner-Dyson distribution of energy level intervals \cite{Wigner:1951,Wigner:1955,Wigner:1957,Wigner:1958,Dyson:1:1962,Dyson:2:1962,Dyson:3:1962}, Porter-Thomas statistics of resonance widths \cite{Porter:1956} and Ericson fluctuations \cite{Ericson:1960,Ericson:1963}. Experimental ultracold molecular samples posses a purity and precision control over all internal and external degrees of freedom at the level of single quantum states which combined with the high DOS makes them the perfect system to make such statistical measurements of chaos. 

In this work we have seen chaos in the spatial degrees of freedom among three atoms.  However, in ultracold collisions of sufficiently anisotropic atoms it is possible that chaotic scattering  may emerge. Indeed, the very recently observed Fano-Feshbach resonances in erbium have exhibited nearest-neighbor statistics corresponding to the Gaussian Orthogonal Ensemble, regarded as a signature of quantum chaos \cite{Frisch:Choas:2013}.  Chaos also affords a new theoretical perspective on cold and ultracold molecular collisions offering the prospect provides deep and unifying insight \cite{Flambaum:2006,Mayle:2012,Mayle:2013}.

\section{Acknowledgments}
This work was supported by the Air Force Office of Scientific Research under the Multidisciplinary University Research Initiative Grant No. FA9550-1-0588. We acknowledge useful discussions with Michael Mayle.

\bibliography{jmh_all,jfec_all}

\end{document}